# Interpreting Cosmological Information from Neural Networks in the Hydrodynamic Universe

Arnab Lahiry[1,2], Adrian E. Bayer[3,4], and Francisco Villaescusa-Navarro[3,4]
[1] Foundation for Research and Technology—Hellas, 100 N. Plastira, Vassilika Vouton, Heraklion, Crete 700 13, Greece; arnablahiry08@gmail.com
[2] Department of Physics, University of Crete, Voutes University Campus, Heraklion, Crete 700 13, Greece
[3] Department of Astrophysical Sciences, Princeton University, Peyton Hall, Princeton, NJ 08544, USA
[4] Center for Computational Astrophysics, Flatiron Institute, 162 5th Avenue, New York, NY 10010, USA

## Abstract

What happens when a black box (neural network) meets a black box (simulation of the Universe)? Recent work has shown that convolutional neural networks (CNNs) can infer cosmological parameters from the matter density field in the presence of complex baryonic processes. A key question that arises is, which parts of the cosmic web is the neural network obtaining information from? We shed light on the matter by identifying the Fourier scales, density scales, and morphological features of the cosmic web that CNNs pay most attention to. We find that CNNs extract cosmological information from both high- and low-density regions: overdense regions provide the most information per pixel, while underdense regions—particularly deep voids and their surroundings—contribute significantly owing to their large spatial extent and coherent spatial features. Remarkably, we demonstrate that there is negligible degradation in cosmological constraining power after aggressive cutting in both maximum Fourier scale and density. Furthermore, we find similar results when considering both hydrodynamic and gravity-only simulations, implying that neural networks can marginalize over baryonic effects with minimal loss in cosmological constraining power. Our findings point to practical strategies for optimal and robust field-level cosmological inference in the presence of uncertainly modeled astrophysics.

## 1. Introduction

Machine learning methods have become a fast-growing tool for tackling various fundamental problems in astrophysics. One such problem is inferring cosmological parameters from the non-Gaussian distribution of matter in the cosmic web, particularly in the presence of baryonic processes whose physics is poorly understood.

For quantities described by Gaussian density fields, such as temperature fluctuations in the cosmic microwave background (CMB), analytical summary statistics, such as the power spectrum, can completely capture the statistical properties of the field. However, the nonlinear evolution of the Universe and the onset of astrophysical processes have transformed the Universe into a non-Gaussian density field, and thus summary statistics are no longer theoretically guaranteed to maximally extract cosmological information. This is relevant for late-time cosmological observables, such as galaxy clustering, weak lensing, 21 cm, and CMB secondaries.

The optimal approach to extract information from such non-Gaussian fields is to perform field-level inference, where every pixel in the map of the density field is used to extract information (see, e.g., J. Jasche & B. D. Wandelt 2013; U. Seljak et al. 2017; J. Jasche & G. Lavaux 2019; F. Schmidt et al. 2019, 2020; T.-X. Mao et al. 2020; N.-M. Nguyen et al. 2021, 2024; B. Dai & U. Seljak 2022, 2024; A. E. Bayer et al. 2023a, 2023b; A. Kostić et al. 2023; A. Andrews et al. 2024; H. Jia 2024; P. Lemos et al. 2024; D. Sharma et al. 2024; B. Horowitz & Z. Lukic 2025). In particular, F. Villaescusa-Navarro et al. (2021b) performed simulation-based field-level inference using convolutional neural networks (CNNs) on 2D projected maps of the total matter density field in hydrodynamic simulations. Their results demonstrated that CNNs can successfully constrain $\Omega_m$ and $\sigma_8$ even when marginalizing over astrophysical uncertainties. While this shows that neural networks can extract cosmological information from the hydrodynamic Universe, it raises a crucial question: what structures in the cosmic web are driving these predictions?

In this work, we aim to interpret the CNN predictions from F. Villaescusa-Navarro et al. (2021b), identifying the morphological features, scales, and density regimes of the cosmic web that contribute most strongly to cosmological inference, paying particular consideration to the impact of baryonic effects on the inference. Not only does this provide insight into the inner workings of neural networks, but it also provides a motivation for how to cut, or filter, data to ensure optimal and robust information extraction from cosmological surveys. It also sheds light on the relative information in regions of high density (e.g., halos) and low density (e.g., voids; C. D. Kreisch et al. 2020; A. E. Bayer et al. 2021, 2022, 2024; C. D. Kreisch et al. 2021; L. Thiele et al. 2023; M. Golshan & A. E. Bayer 2024), or more generally, the information content of the cosmic web (J. Sunseri et al. 2025).

Other works toward interpreting neural networks in different astrophysical contexts include L. Lucie-Smith et al. (2018, 2019, 2024a, 2024b), J. M. Zorrilla Matilla et al. (2020), P. Bhambra et al. (2022), W. You et al. (2023), H. Chen et al. (2024), and M. Golshan & A. E. Bayer (2024).

The paper is structured as follows: We discuss our method in Section 2. We then present our results in Section 3. Finally, we conclude in Section 4.

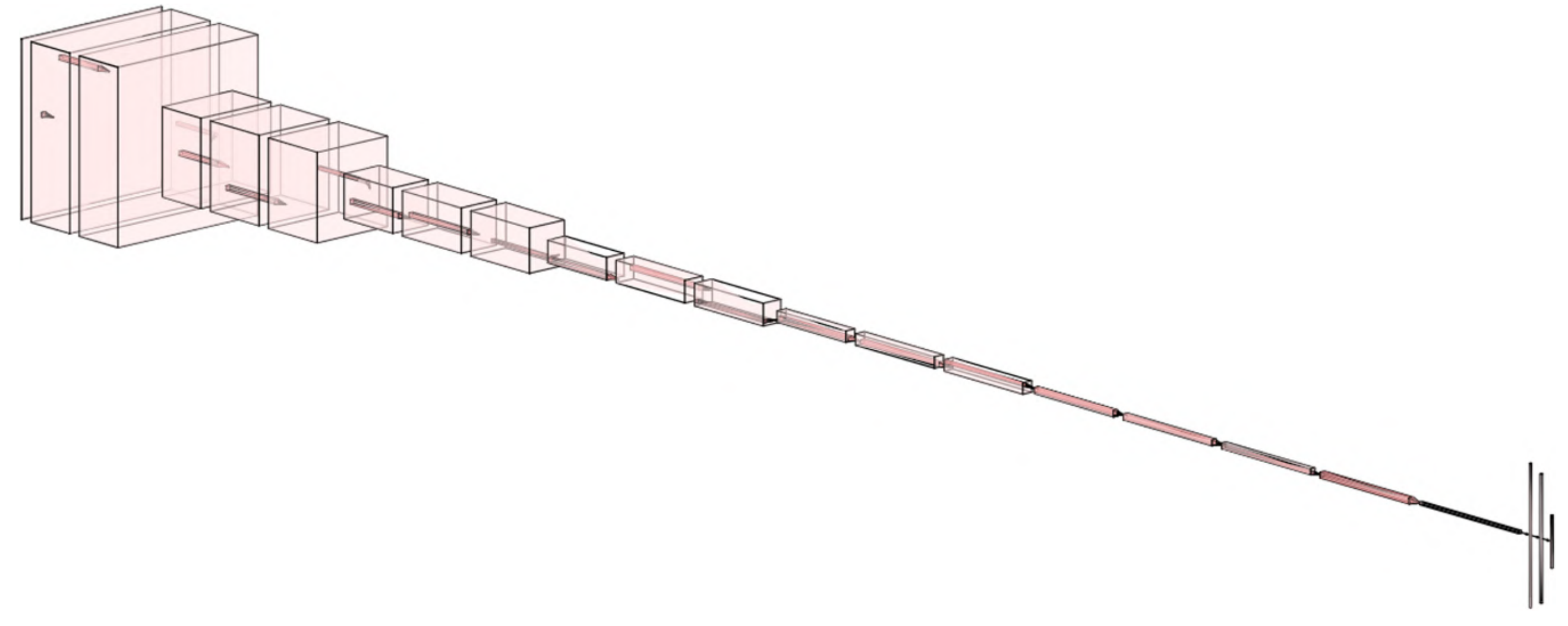

**Figure 1.** Schematic of the CNN architecture employed in parameter estimation. There are sets of multiple convolution layers with a kernel size = 4, stride = 2, and padding = 1, coupled with 2D batch normalization and `LeakyReLU` activation functions. The 256 × 256 × 1 image is convolved and flattened into a 1D array and is then connected to an output array of size 12 (posterior mean and variance for the six inferred parameters) using fully connected layers.

**Table 1**
Description of the Cosmological and Astrophysical Parameters for Each Simulation Suite and the Numerical Range of Variation in CAMELS

| Parameter | Description | Range |
|---|---|---|
| $\Omega_m$ | Energy density of total matter (dark + baryonic) as compared to the critical density of the Universe | [0.1, 0.5] |
| $\sigma_8$ | Variance of the power spectrum of the linear field at redshift = 0, at a radial distance of 8 $h^{-1}$ Mpc | [0.6, 1.0] |
| $A_{\rm SN1}$ | Galactic winds due to supernova feedback: energy per unit star formation rate | [0.25, 4.0] |
| $A_{\rm AGN1}$ | Kinetic mode black hole feedback, representing the energy per unit black hole accretion rate | [0.25, 4.0] |
| $A_{\rm SN2}$ | Wind speed of the galactic winds caused by supernova feedback | [0.5, 2.0] |
| $A_{\rm AGN2}$ | Kinetic mode black hole feedback, representing matter ejection speed | [0.5, 2.0] |

## 2. Methods

In this section, we describe the data in Section 2.1, the neural network in Section 2.2, and the interpretability analysis methods in Section 2.3, and we overview the analysis pipeline in Section 2.4.

### 2.1. Simulations

In this work, we use data from the IllustrisTNG Latin Hypercube (LH) set of the CAMELS project (F. Villaescusa--Navarro et al. 2021a). CAMELS comprises over 16,000 simulations, both (magneto)hydrodynamic and gravity-only, spanning a wide range of two cosmological parameters ($\Omega_m$, $\sigma_8$) and four astrophysical feedback parameters ($A_{\rm SN1}$, $A_{\rm SN2}$, $A_{\rm AGN1}$, $A_{\rm AGN2}$). The simulations of the IllustrisTNG suite have been run with the AREPO code (R. Weinberger et al. 2019) and employ the same subgrid physics model as the original IllustrisTNG simulations (A. Pillepich et al. 2018; D. Nelson et al. 2019).

All simulations follow the evolution of $2 \times 256^3$ dark matter and fluid elements in a periodic comoving volume of $(25\ h^{-1}\ {\rm Mpc})^3$ from $z = 127$ down to $z = 0$. The astrophysical effects are parameterized by four parameters that control the efficiency of supernova and active galactic nucleus (AGN) feedback: $A_{\rm SN1}$, $A_{\rm SN2}$, $A_{\rm AGN1}$, and $A_{\rm AGN2}$. A short description of the physical meaning of each parameter can be found in Table 1. In the LH set, each simulation has a different value of the astrophysical parameters, as well as different values of the cosmological parameters $\Omega_m$ and $\sigma_8$. Other cosmological parameters are fixed as follows: $\Omega_b = 0.049$, $h = 0.6711$, $n_s = 0.9624$, $\sum m_\nu = 0.0$ eV, $w = -1$. Each simulation also has a different value for the initial random seed.

The simulation snapshots from the IllustrisTNG LH set are used to render 2D maps of the total matter field $M_{\rm tot}$ at $z = 0$ with dimensions (256 × 256). These maps are available in the CAMELS Multifield Data set (CMD; F. Villaescusa-Navarro et al. 2022); we refer the reader to F. Villaescusa-Navarro et al. (2021a) and F. Villaescusa-Navarro et al. (2022) for further details on the simulations of the CAMELS project.

### 2.2. Neural Network

To interpret the cosmological results obtained by F. Villaescusa-Navarro et al. (2021b), we mimic the CNN architecture of F. Villaescusa-Navarro et al. (2021b). The architecture consists of a set of convolutional layers, combined with `batchnorm` and `LeakyReLU` nonlinear activation layers, followed by two fully connected layers—a rough schematic is shown in Figure 1. The CNN was used to perform simulation-based inference to find the mapping between the 2D images and the posterior, denoted

$$p(\boldsymbol{\theta}|\boldsymbol{X}), \tag{1}$$

where $\boldsymbol{X}$ is the 2D map and $\boldsymbol{\theta}$ represents the six simulation parameters ($\Omega_m$, $\sigma_8$, $A_{\rm SN1}$, $A_{\rm AGN1}$, $A_{\rm SN2}$, $A_{\rm AGN2}$). The model predicts the mean and variance of the marginal posterior of all six parameters:

$$\mu_i(\boldsymbol{X}) = \int_{\theta_i} p(\theta_i|\boldsymbol{X})\theta_i d\theta_i \tag{2}$$

$$\sigma_i^2(X) = \int_{\theta_i} p(\theta_i|\boldsymbol{X})(\theta_i - \mu_i)^2 d\theta_i, \tag{3}$$

where $\mu_i$, $\sigma_i^2$, and $\theta_i$ are the predicted mean, predicted variance, and true value of the $i$th parameter, respectively, and $p(\theta_i|\boldsymbol{X})$ is the marginal posterior over the $i$th parameter. This is achieved by minimizing the following loss function (N. Jeffrey & B. D. Wandelt 2020; F. Villaescusa-Navarro et al. 2021b):

$$\mathcal{L} = \sum_{i=1}^{6} \log\left(\sum_{j\in \rm batch} (\theta_{i,j} - \mu_{i,j})^2\right) + \sum_{i=1}^{6} \log\left(\sum_{j\in \rm batch} ((\theta_{i,j} - \mu_{i,j})^2 - \sigma_{i,j}^2)^2\right). \tag{4}$$

We use the optuna package (T. Akiba et al. 2019) to perform hyperparameter optimization. The neural network hyperparameters we consider are (1) the number of convolutional layers, (2) the dropout rate for the fully connected layers, (3) the weight decay, and (4) the learning rate. The hyperparameter space is sampled using the TPE sampling method, and 50 trials are conducted to find the optimal set of hyperparameters that best minimize the validation loss.

The data are split into three different sets, training, validation, and testing in the ratio 90:5:5, such that each 2D map comes from a different original simulation box. During training, we perform data augmentation through rotations and flips to enforce invariance to these symmetries.

We quantify the accuracy of the network using the $R^2$ score of the predicted parameters from the test data set, where

$$R^2 = 1 - \frac{\sum_{i=1}^{n}(\theta_i - \mu_i)^2}{\sum_{i=1}^{n}(\theta_i - \bar{\theta})^2}. \tag{5}$$

### 2.3. Interpretability

We consider three different interpretability metrics: Saliency Maps, Integrated Gradients, and GradientSHAP. Each metric interprets the neural network predictions by calculating the contribution of each input feature (in this case, each pixel in the image) to each output feature (the cosmological parameters). We now describe each metric.

#### 2.3.1. Saliency Maps

Saliency Maps (K. Simonyan et al. 2013) provide a measure of how sensitive a model's output is to small changes in the input. Given a model $f$: $\mathbb{R}^n \to \mathbb{R}$ and an input $x \in \mathbb{R}^n$, the Saliency Map is defined as the gradient of the model's output with respect to the input

$$S_i(x) = \frac{\partial f(x)}{\partial x_i}, \tag{6}$$

where $S_i(x)$ measures how much a small change in the $i$th input feature would affect the model's output. Large values of $S_i(x)$ indicate that the corresponding feature $x_i$ is important for the model's prediction. Saliency Maps capture *local sensitivity*, but they can be noisy and are prone to gradient saturation in deep networks.

#### 2.3.2. Integrated Gradients

Integrated Gradients (M. Sundararajan et al. 2017) address the problem of saturation and local noise by computing the contribution of each feature along a straight path from a baseline input $x'$ (often set to zero) to the actual input $x$. The Integrated Gradients for the $i$th input feature is defined as

$$\mathrm{IG}_i(x) = (x_i - x_i') \int_0^1 \frac{\partial f(x' + \alpha(x - x'))}{\partial x_i} d\alpha, \tag{7}$$

where $x'$ is the baseline input (often white noise or a blank image), $\alpha$ interpolates between the baseline and the actual input, and the integral averages the gradients along the path from $x'$ to $x$.

Integrated Gradients provide a more stable and *global* measure of feature importance compared to raw gradients.

#### 2.3.3. GradientSHAP

GradientSHAP (S. Lundberg & S.-I. Lee 2017) extends Integrated Gradients by incorporating the idea of Shapley values to account for feature interactions and uncertainty. It computes the expected gradients with respect to a distribution of baselines $p(x')$ to account for variability in the attributions

$$\phi_i(x) = \mathbb{E}_{x'\sim p(x')}\left[(x_i - x_i') \int_0^1 \frac{\partial f(x' + \alpha(x - x'))}{\partial x_i} d\alpha\right], \tag{8}$$

where $p(x')$ is the distribution over possible baseline inputs. GradientSHAP computes the average Integrated Gradients over multiple baselines, capturing the effect of feature interactions and uncertainty.

#### 2.3.4. Interpretability Summary

In summary, Saliency Maps capture local sensitivity but can be noisy and suffer from gradient saturation, Integrated Gradients provide a global measure of feature importance along a path from a baseline to the input, and GradientSHAP generalizes Integrated Gradients by averaging over a distribution of baselines, improving stability and accounting for feature interactions. We use the `Captum` library (N. Kokhlikyan et al. 2020) in `Python` to implement the algorithms.

### 2.4. Analysis Pipeline

To interpret how cosmological information is extracted by the neural network, we perform various experiments, cutting out information from regions of the map with different Fourier scales $k$ and densities $\rho$. We implement an end-to-end analysis pipeline comprising four major stages:

1. *Preprocessing and Cutting of Input Maps.* We begin by selecting 2D total matter density maps from the CAMELS simulations. These maps are filtered using specific criteria, such as maximum Fourier scale cuts ($k_{\rm max}$) or density cuts ($\rho_{\rm min}$, $\rho_{\rm max}$), in order to isolate distinct regimes to focus our analysis (e.g., voids or halos). For $k$ cuts this is achieved with a top-hat filter, while for $\rho$ cuts this is performed by replacing the pixels outside of the cut threshold by a random number between the global minimum and maximum of the pixel values, distributed uniformly in $\log\rho$.

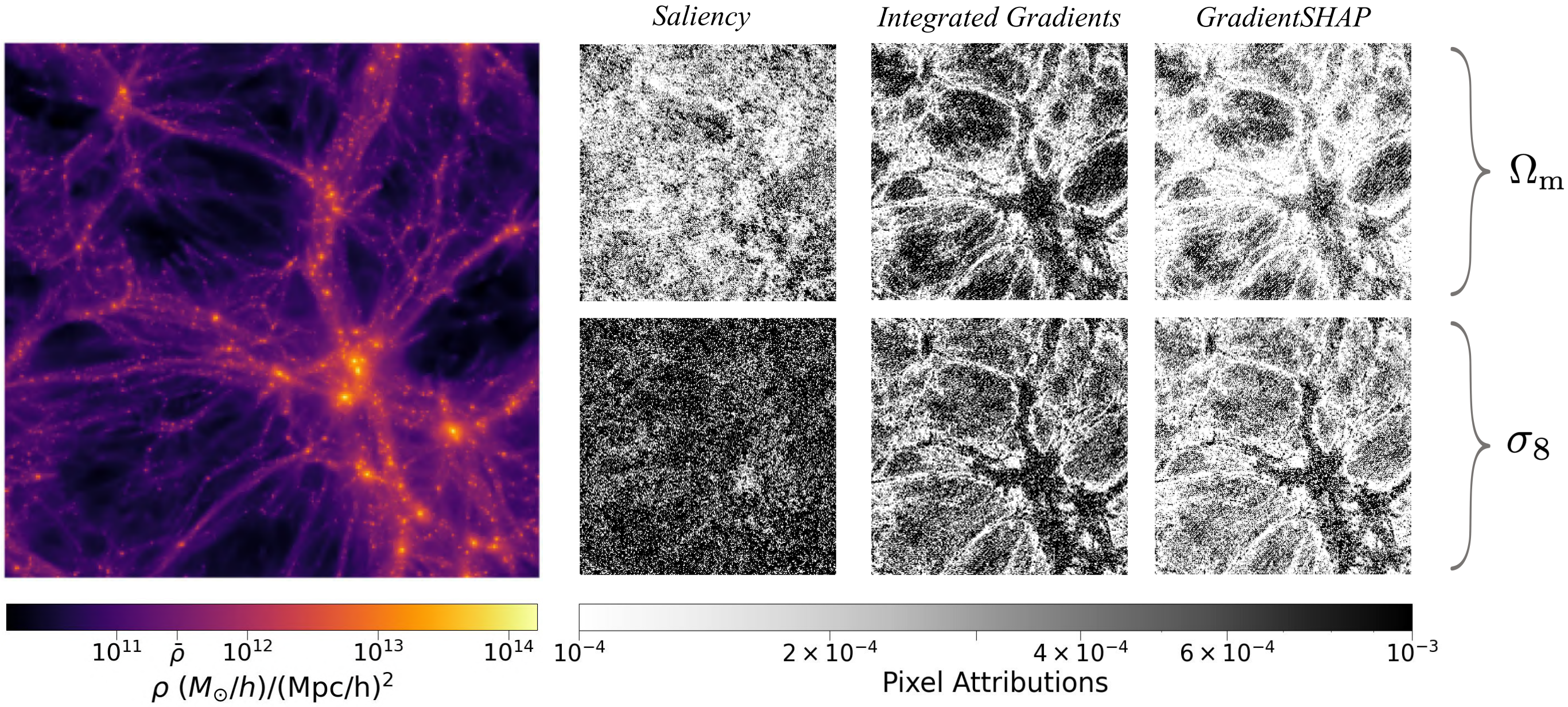

**Figure 2.** Interpretability maps. Saliency Map, Integrated Gradients, and GradientSHAP interpretability methods for $\Omega_m$ (top right) and $\sigma_8$ (bottom right) for an example image of the cosmic web (left). The cosmic web map is scaled logarithmically. Black color in the attribution plots refers to features that are important to the neural network, while white implies low importance.

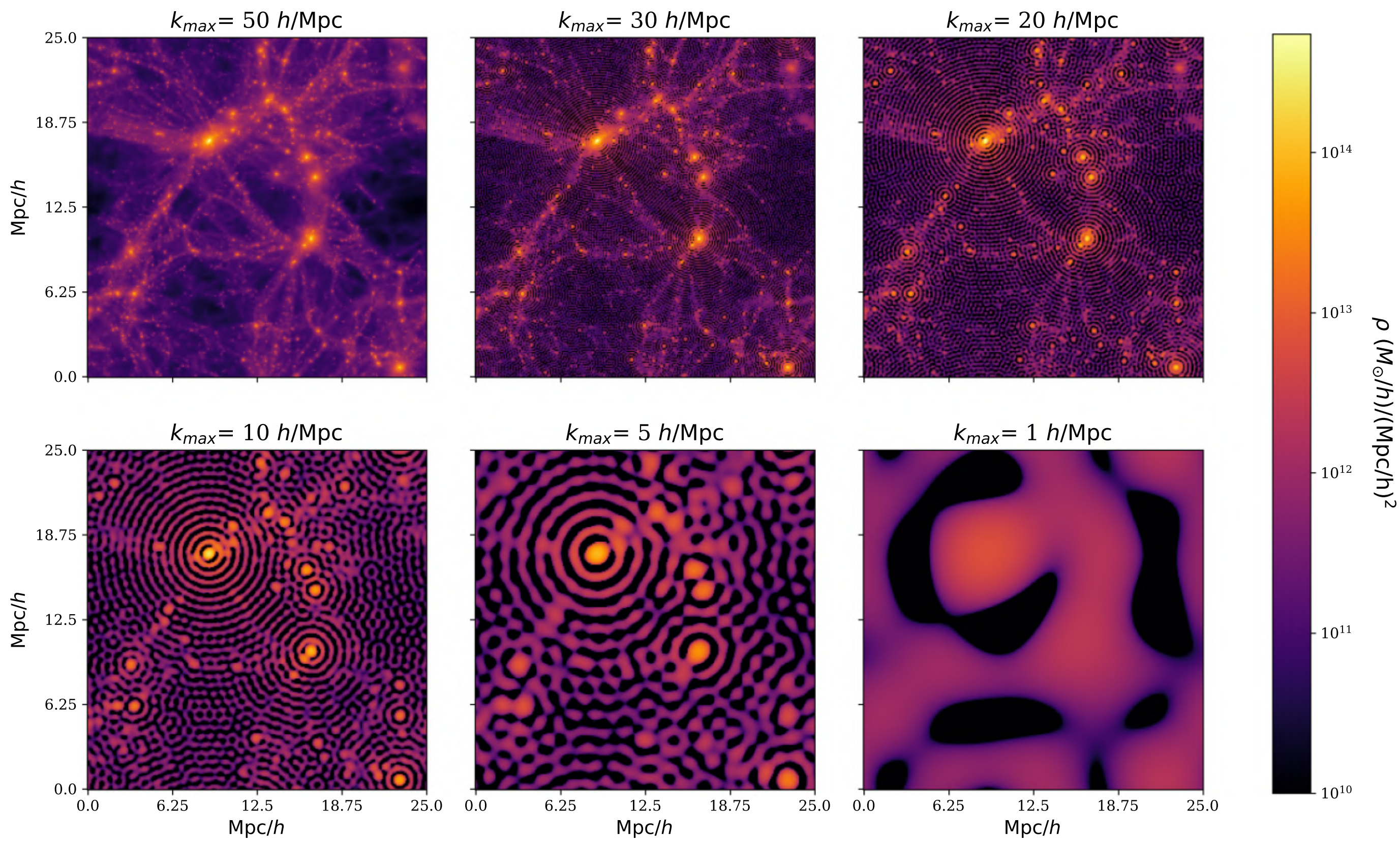

**Figure 3.** Fourier scale cuts. An example of applying a top-hat filter in $k$-space to a total matter density map for $k_{\rm max} = 50$, 30, 20, 10, 5, and 1 $h/\mathrm{Mpc}$.

2. *Training the Neural Network.* For each experiment, the maps are divided into training, validation, and testing sets in the ratio 90:5:5. A CNN is trained to perform simulation-based inference and infer the two cosmological parameters $\Omega_m$ and $\sigma_8$, as well as four astrophysical parameters. The network is optimized to predict both the mean and variance of the marginal posterior for each parameter by minimizing a custom loss function.

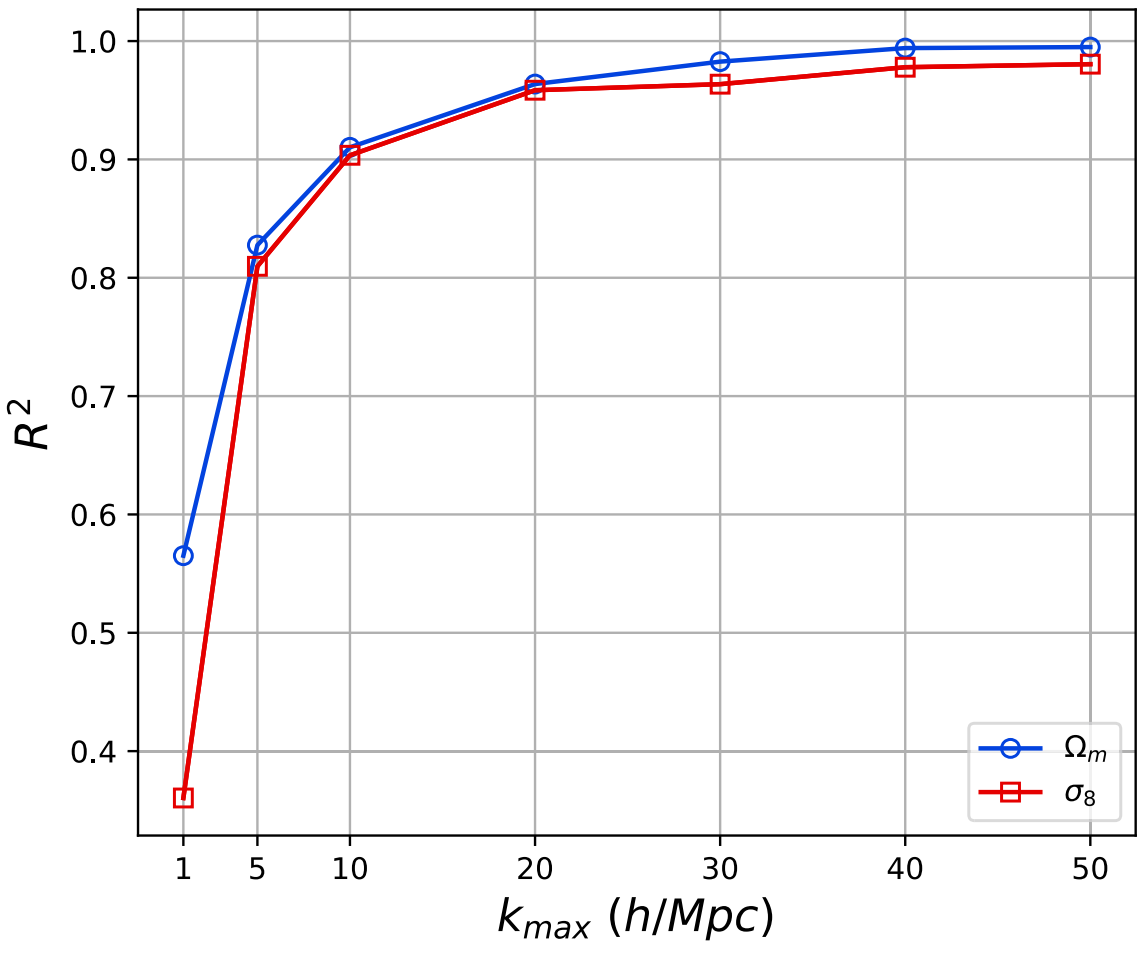

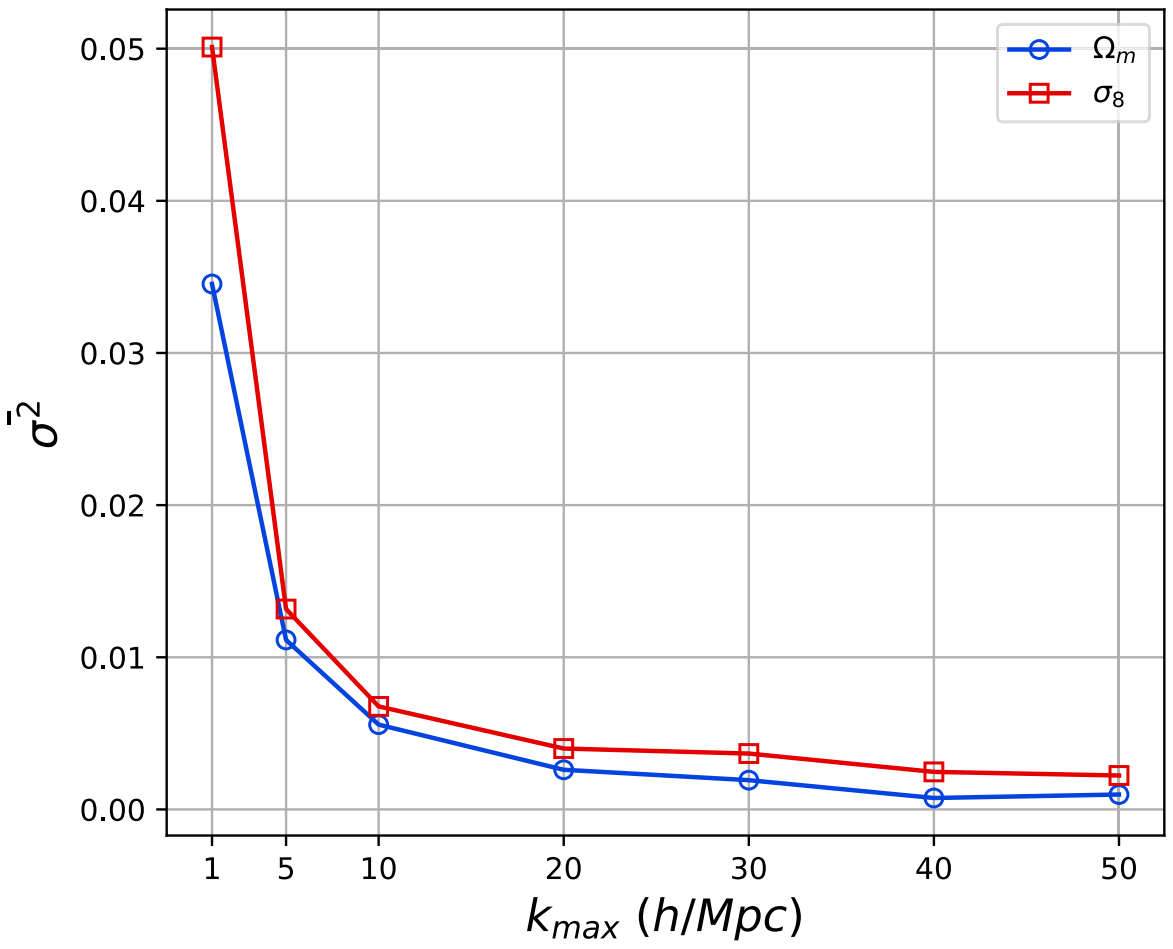

**Figure 4.** Fourier scale cut inference. The $R^2$ scores (left) and average variance $\bar{\sigma}^2$ (right) of the predicted cosmological parameters $\Omega_m$ (circles in blue) and $\sigma_8$ (squares in red) as a function of $k_{\max}$ for the hydrodynamic maps. Note that the Nyquist frequency is $\sim$32 $h$/Mpc. At low $k_{\max}$ there is a sharp increase in the $R^2$ and a decrease in variance up to around $k_{\max} \sim 20\, h$/Mpc, with only a slight improvement when including smaller scales.

3. *Interpretability and Attribution Mapping.* Upon completion of training, we apply the Integrated Gradients interpretability algorithm to the test data for the inference of the cosmological parameters. This method generates attribution maps that quantify the contribution of each input pixel to the network's prediction, enabling us to identify which structures in the input maps (e.g., voids, filaments, halos) are most informative.
4. *Cosmological Parameter Constraints.* We then assess the neural network's predictive performance by computing the coefficient of determination ($R^2$) and the inferred variance ($\sigma^2$). These statistics provide case-wise insights into the accuracy and precision of the network for the different $k$ and $\rho$ cuts performed, allowing us to interpret the physical origin of the extracted cosmological information and to quantify the relative importance of different regions of the cosmic web.

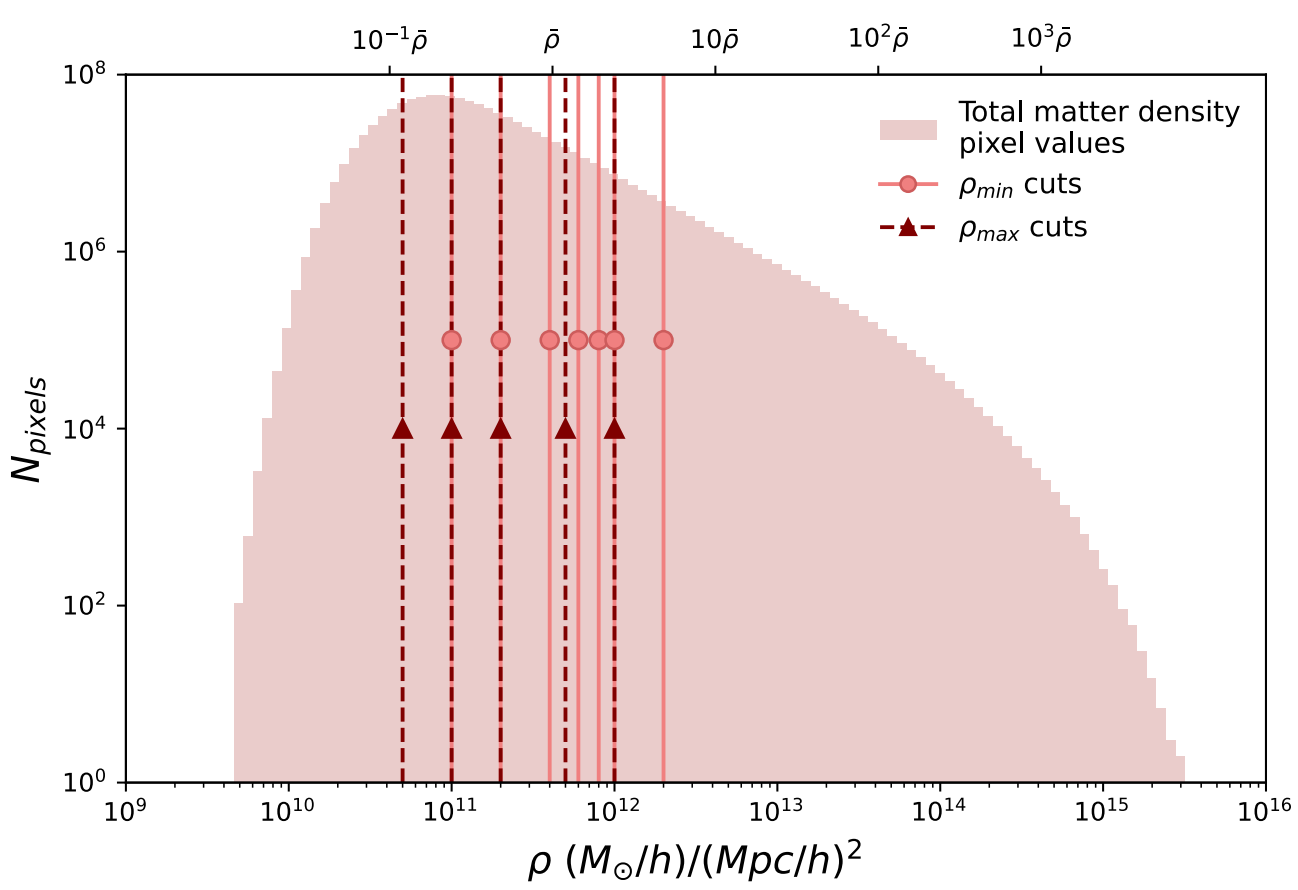

**Figure 5.** Histogram of the matter density pixel values. The dashed lines (triangles) represent the upper-limit density cuts ($\rho_{\max}$), and the solid lines (circles) represent the lower-limit density cuts ($\rho_{\min}$). The top axis shows the values of $\rho$ in various multiples of the average density $\bar{\rho} = 4.29 \times 10^{11}$ $(M_\odot/h)/(\mathrm{Mpc}/h)^2$ at $z = 0$.

## 3. Results

We now interpret the source of cosmological information for the complete set of uncut maps in Section 3.1. We then perform Fourier scale cuts in Section 3.2 and density cuts in Section 3.3.

### 3.1. Where Is the Information?

We start by considering the maps in their entirety, taking no cuts. We use the Saliency Map, Integrated Gradients, and GradientSHAP methods described in Section 2.3 to calculate attributions for $\Omega_m$ and $\sigma_8$, marginalizing over the astrophysical parameters. Figure 2 shows the results, with black representing pixels with a higher attribution, indicating areas with greater sensitivity to changes in input. We see a great similarity between Integrated Gradients and GradientSHAP, which both consider global features, while the Saliency Map results are less conclusive, in particular, because they only look at local behavior and are prone to noise due to gradient saturation.

Upon comparing the interpretability maps with the input density fields, we find that for $\Omega_m$ the CNN pays most attention to pixels within deep void regions and clusters, implying that the CNN is learning from the most extreme regions in the map. Interestingly, the thick filaments have relatively high attribution, while the thin filaments have relatively little attribution. We find similar results for $\sigma_8$, with the difference that now overdense regions have slightly higher attributions than underdense regions.

In this work, we are interested in the global features of the cosmic web and thus focus on the Integrated Gradients approach for the remainder of the paper, particularly because we find little difference between the overall features of the Integrated Gradients and GradientSHAP maps. In addition, the Integrated Gradients method balances interpretability, mathematical robustness, and computational expense best among the three algorithms.

### 3.2. Fourier Scale Analysis

We first investigate how the model's accuracy depends on the maximum Fourier scale $k_{\max}$. On the one hand, the number of modes goes as $k_{\max}^2$ (for 2D fields), and therefore one may expect that models trained on higher-resolution images will yield more precise inferences. On the other hand, baryonic effects contaminate the data on small scales, making it more difficult to infer cosmological parameters. To quantify the trade-off of these effects, we start by convolving the maps with a top-hat $k$ filter with various $k_{\mathrm{max}}$ values. This procedure will set any mode with $k \geqslant k_{\mathrm{max}}$ to zero. We apply no $k_{\mathrm{min}}$ cut. Figure 3 shows the effect of this filtering on an example map. The filtered maps are then passed through the CNN.

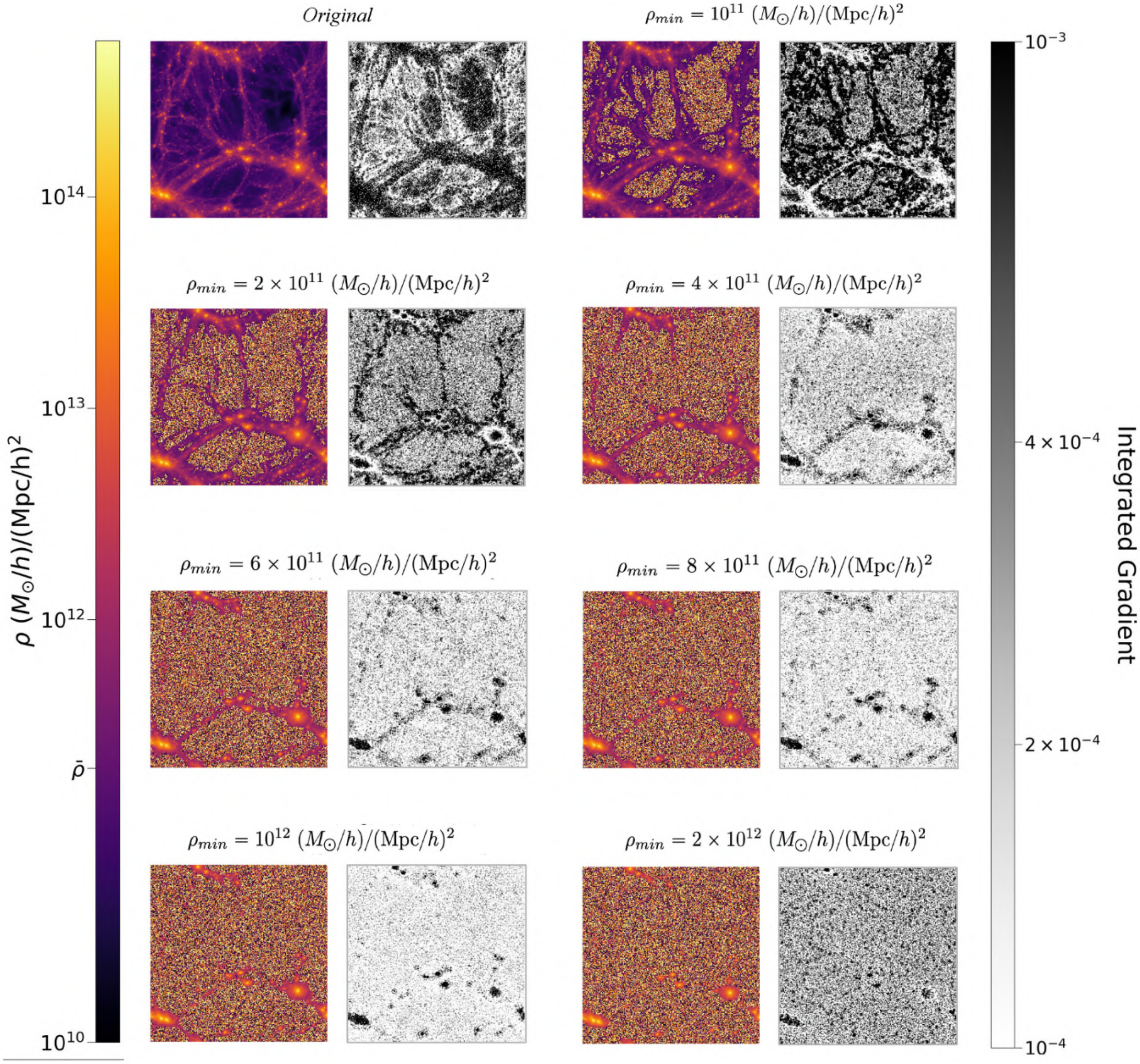

**Figure 6.** Underdensity cut Integrated Gradients. Illustration of the underdensity cutting procedure applied to density maps. For each $\rho_{\mathrm{min}} = 10^{11}$, $2 \times 10^{11}$, $4 \times 10^{11}$, $6 \times 10^{11}$, $8 \times 10^{11}$, $10^{12}$, and $2 \times 10^{12}$ $(M_\odot/h)/(\mathrm{Mpc}/h)^2$, pixel values below $\rho_{\mathrm{min}}$ are replaced with random values uniformly sampled from the global minimum and maximum density values across all maps. These augmented maps (left columns) are used to train the neural network to assess the impact of underdensity cutting on cosmological parameter inference. The right column shows the corresponding Integrated Gradients attribution maps for $\Omega_m$, with Integrated Gradients values normalized between 0 and 1 (the black-and-white scale is chosen for optimal visualization of patterns).

Figure 4 shows the $R^2$ score (left) and average variance predicted via simulation-based inference (right) for $\Omega_m$ and $\sigma_8$ as a function of $k_{\mathrm{max}}$ (see Appendix A.3 for explicit truth vs. prediction plots for different levels of $R^2$, for intuition). The network is able to obtain $R^2 \sim 1$ on the cosmological parameters even in the presence of baryonic effects. We observe that when going to higher values of $k_{\mathrm{max}}$ there is a sharp increase in the $R^2$ and a decrease in variance up to around $k_{\mathrm{max}} \sim 20\,h/\mathrm{Mpc}$, with only a slight improvement when including smaller scales. This implies that cutting out the smallest scales, which suffer from the highest baryonic contamination, and which may not be well understood when confronting real data, could be safely removed without destroying much cosmological information. The low $R^2$ at $k \sim 1\,h/\mathrm{Mpc}$ is somewhat related to the small box size of the CAMELS simulations—for larger boxes, the $k$ at which $R^2$ becomes approximately unity will be lower.

### 3.3. Density Cut Analysis

Having found that the neural network is sensitive to different components of the cosmic web, we perform a more quantitative analysis by considering the performance of the network as a function of different density cuts. In particular, we cut out information in high- and low-density regions and investigate the performance of the network in both cases. Figure 5 shows a histogram of the mass density $\rho$ in the pixels from all 15,000 total matter density maps. The lowest values of $\rho$ represent deep voids in the cosmic web, and the higher pixel values represent clustered matter such as cosmic filaments, with the highest values representing halos. Roughly speaking, in terms of the average density $\bar{\rho} = 4.29 \times 10^{11}$ $(M_{\odot}/h)$ $({\rm Mpc}/h)^2$, voids correspond to $\rho \lesssim \bar{\rho}/2$, filaments correspond to $5\bar{\rho} \lesssim \rho \lesssim 50\bar{\rho}$, and halos correspond to $\rho \gtrsim 10^2\bar{\rho}$.

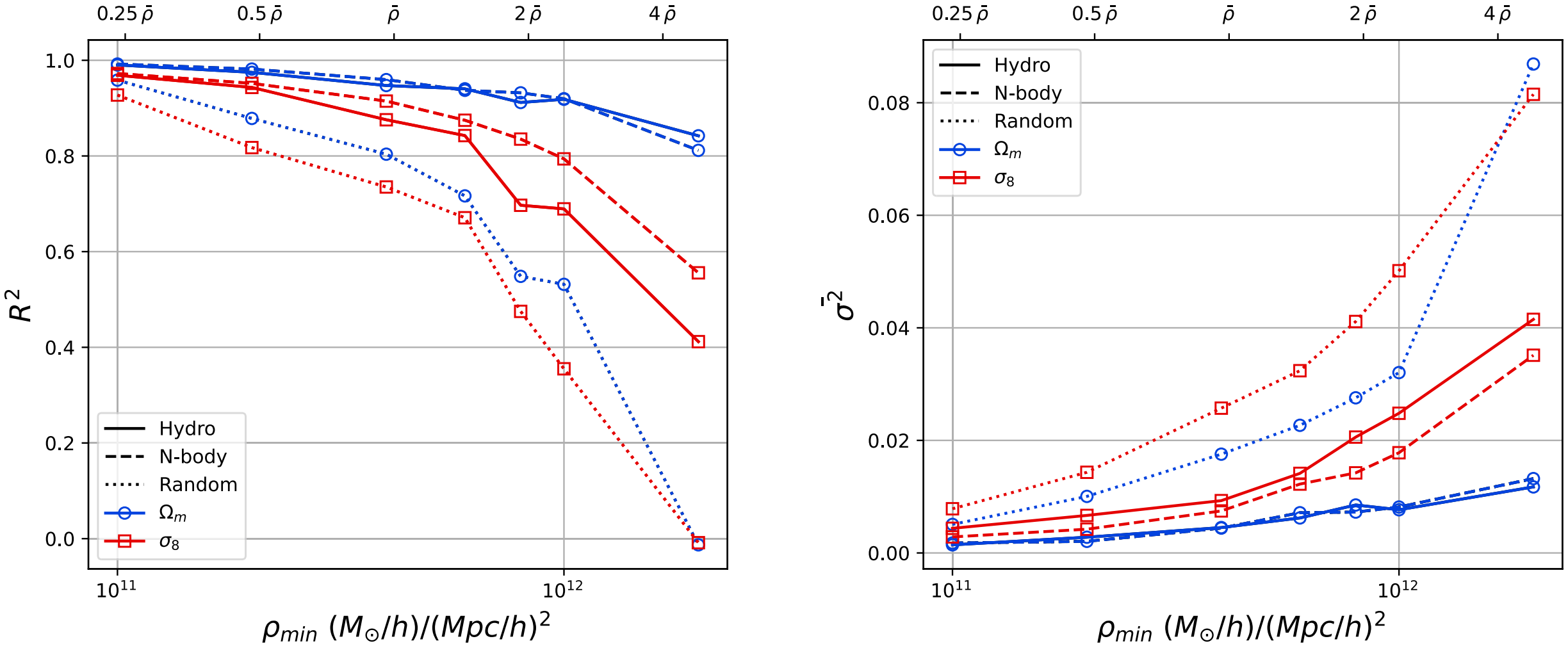

**Figure 7.** Underdensity cut inference. The $R^2$ scores (left) and average variance $\bar{\sigma}^2$ (right) of the predicted cosmological parameters $\Omega_m$ (circles in blue) and $\sigma_8$ (squares in red) as a function of $\rho_{\rm min}$. The solid lines represent the maps from the hydrodynamic simulations, the dashed lines represent the maps from the $N$-body simulations, and the dotted line represents the case where an equal number of pixels to the corresponding $\rho_{\rm min}$ case is randomly augmented for each $N$-body map.

#### 3.3.1. Underdensity Cuts

We start by considering underdensity cuts, by removing information from regions with density less than $\rho_{\rm min}$. Every pixel that has a density below $\rho_{\rm min}$ is replaced by a random number between the global minimum and maximum pixel values, which are uniformly sampled on a logarithmic scale. This method essentially "deletes" any information below $\rho_{\rm min}$ by replacing it with uniform noise. Figure 5 shows the values of the $\rho_{\rm min}$ cuts, which run from $10^{11}$ to $2 \times 10^{12}$ $(M_{\odot}/h)/$ $({\rm Mpc}/h)^2$ in multiples of 2. We retrain the CNN to infer $\Omega_m$ and $\sigma_8$ for each different $\rho_{\rm min}$ cut.

Figure 6 depicts the gradual cutting of the pixels for the different minimum density cuts, as well as the Integrated Gradients maps for the prediction of $\Omega_m$. It can be seen that when there is no density cut the Integrated Gradients are large at both deeply underdense and highly overdense regions, but as the minimum density cut is applied, the network can no longer learn anything from underdense regions.

For a more quantitative analysis, Figure 7 shows the change in $R^2$ and $\bar{\sigma}^2$ for $\Omega_m$ and $\sigma_8$ as a function of $\rho_{\rm min}$. First focusing on the solid lines, which are for the full hydrodynamical CAMELS simulations, it can be seen that $R^2$ is approximately 1 for the lowest $\rho_{\rm min}$, gradually decreasing as $\rho_{\rm min}$ is increased. Similarly, the change in $\bar{\sigma}^2$ is relatively small at low $\rho_{\rm min}$. For the highest cut we consider, $\rho_{\rm min} \sim 5\bar{\rho}$, the $R^2$ has dropped to approximately 0.8 for $\Omega_m$ and 0.6 for $\sigma_8$. This implies that much information is lost by removing underdense regions, particularly voids and filaments. However, it is important to note that the network may still be able to infer the presence of voids from coherent noisy structures in the maps, even when the most underdense regions are cut. This suggests that while there is some information loss, it is not dramatic. The model's ability to recognize voids might stem from the spatial coherence of these regions, which are distinguishable from the surrounding dense structures. Furthermore, the fact that $R^2$ and $\bar{\sigma}^2$ gradually change for all cuts considered implies that there is information in the deepest (regions of) voids that have the lowest $\rho$ values.

We now perform further analysis to understand what causes the degradation in information for higher $\rho_{\rm min}$. First, to investigate whether this information loss is because of the coherent structures in the dark matter distribution or because high-density regions are more sensitive to astrophysical effects, which we are marginalizing over here, we repeat the analysis using the maps from pure $N$-body simulations. These correspond to gravity-only simulations that only model the evolution of cold dark matter and do not implement astrophysical processes like stellar and black hole feedback. These results are shown by the dashed lines in Figure 7. It can be seen that for $\Omega_m$ there is almost no difference between the inference on full hydro or $N$-body, while for $\sigma_8$ there is some information degradation from astrophysical effects. Nevertheless, whether one uses hydro or $N$-body, the general trends of information loss as a function of $\rho_{\rm min}$ are the same, implying that marginalization over astrophysical effects is not the source of this information degradation.

Next, we investigate whether this information loss is simply due to the number of cut pixels. For this experiment, we randomize the $\rho$ pixel values such that for each $\rho_{\rm min}$ we randomize the same number of pixels as the number of pixels with $\rho < \rho_{\rm min}$. Figure 8 shows the randomly augmented maps and Integrated Gradients maps for the model trained on this data set. Interestingly, it can be seen that the model is now unable to recognize the features describing the higher-density filaments and halos and is extracting the cosmological information from the relatively unobscured pixels depicting the voids. It can be argued that since CNNs rely on local convolutions for feature recognition, arbitrarily removing pixels hinders the ability of the network to extract any information from the finer structures, namely filaments and halos. The dotted lines in Figure 7 show the results for these random augmentations, and it can be seen that the model now performs worse than when cuts are made based on the density. This all implies that the model is learning something from the coherent structures of the matter field, i.e., if one had to cut a given number of pixels in a map, one would lose less information by selecting these pixels based on a physical density cut, instead of randomly cutting.

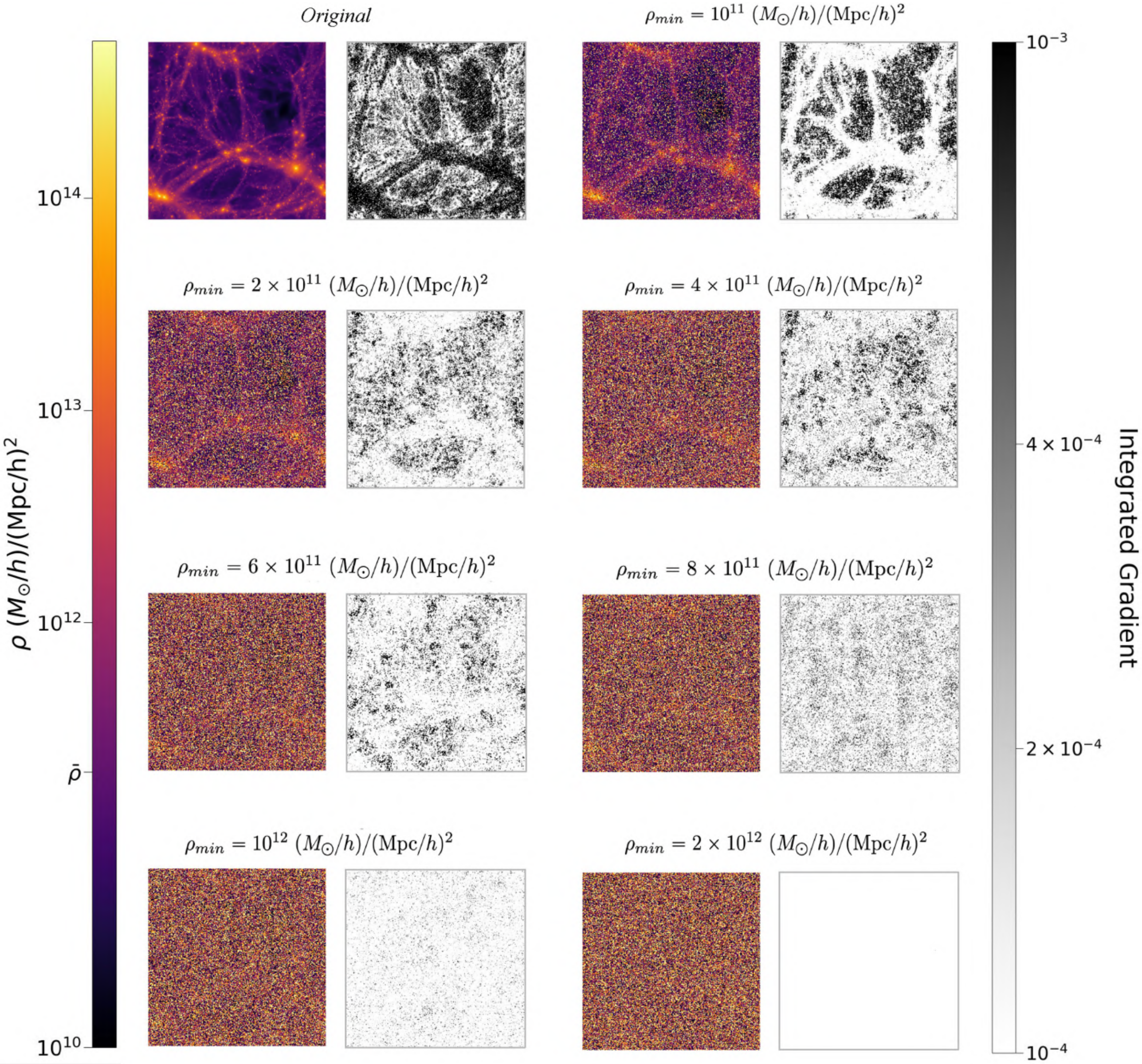

**Figure 8.** Random density cut Integrated Gradients. Illustration of the random cutting procedure where, for each underdensity threshold value $\rho_{\rm min} = 10^{11}$, $2 \times 10^{11}$, $4 \times 10^{11}$, $6 \times 10^{11}$, $8 \times 10^{11}$, $10^{12}$, and $2 \times 10^{12}$ $(M_\odot/h)/(\mathrm{Mpc}/h)^2$, an equal number of pixels are randomly selected and replaced with values uniformly sampled from the global minimum and maximum across all maps. These randomly augmented maps (left column) are used to examine whether the cosmological inference is sensitive to the spatial structure of the removed pixels. The right column shows the corresponding Integrated Gradients attribution maps for $\Omega_m$, with Integrated Gradients values normalized between 0 and 1 (the black-and-white scale is chosen for optimal visualization of patterns).

#### *3.3.2. Overdensity Cuts*

We now consider overdensity cutting, by cutting information from pixels with densities larger than $\rho_{\rm max}$. The values of the cuts can be seen in Figure 5. We choose the lowest $\rho_{\rm max}$ to be approximately $10^{-1}\bar{\rho}$ to consider the internal regions of voids. Similar to the minimum-density analysis case, every pixel that has a density above $\rho_{\rm max}$ is replaced by a uniformly sampled random value in the range of the global minimum and maximum values of all pixels in the logarithmic scale. In addition, as in the previous case, this experiment was done with both hydrodynamical maps and *N*-body maps, but we do not repeat the random pixel augmentation analysis in this case because we expect the conclusions to be similar.

Figure 9 shows the data with different overdensity cuts, as well as the Integrated Gradients maps for $\Omega_m$, and Figure 10 shows how the information varies with $\rho_{\rm max}$. It can be seen that both $R^2$ and $\bar{\sigma}^2$ saturate at $\rho_{\rm max} \sim 2 \times 10^{11}\, M_\odot/h/(\mathrm{Mpc}/h)^2 \sim \bar{\rho}/2$, implying that the network is able to learn almost everything it can about $\Omega_m$ and $\sigma_8$ from relatively underdense void-like regions. It is interesting to note the much more sudden change in $R^2$ and $\bar{\sigma}^2$ for the maximum density cut compared to the minimum density cut analysis in the previous subsection—this sudden change between $\bar{\rho}/10$ and $\bar{\rho}/2$ implies that there is substantial information in shallow voids, in particular, there is a significant change at $\rho_{\rm max} \simeq \bar{\rho}/10$, where most of the information, apart from the deepest voids, has been removed. An alternative explanation is that high-density regions are rare and very localized; thus, even if we make these regions noisy, the noise is coherent in space, and a network may still be able to learn from these regions. In other words, while making a $\rho_{\rm max}$ cut destroys the internal profiles of halos and filaments, the network may be still be using information from the positions of halos, which can still be extracted from the noisy coherent regions.

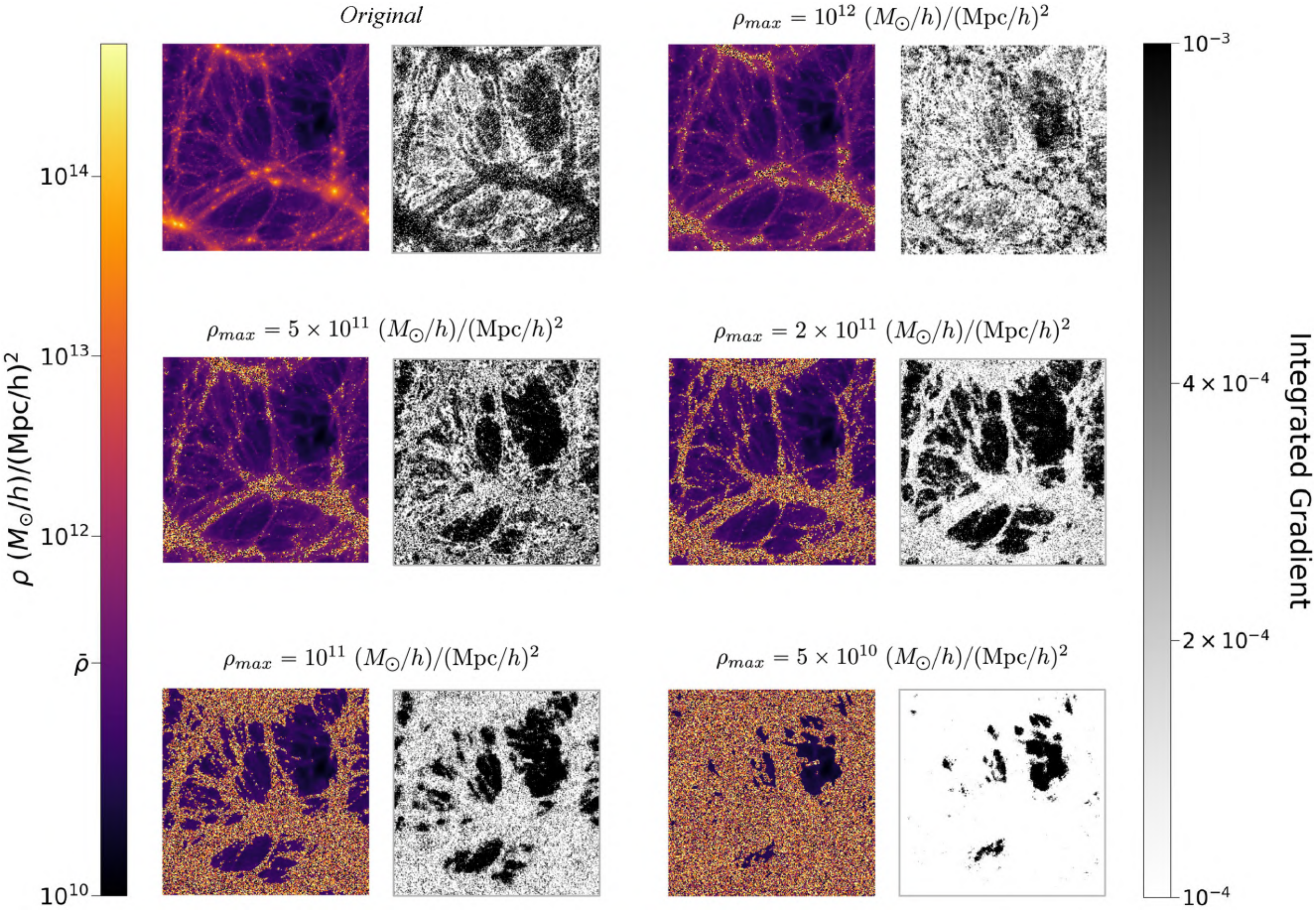

**Figure 9.** Overdensity cut Integrated Gradients. Illustration of the overdensity cutting procedure applied to density maps. For each threshold value $\rho_{\rm max} = 10^{12}$, $5 \times 10^{11}$, $2 \times 10^{11}$, $10^{11}$, and $5 \times 10^{10}$ $(M_\odot/h)/(\mathrm{Mpc}/h)^2$, pixel values above $\rho_{\rm max}$ are replaced with random values uniformly sampled from the global minimum and maximum density values across all maps. These augmented maps (left column) are used to train the neural network to assess the impact of overdensity cutting on cosmological parameter inference. The right column shows the corresponding Integrated Gradients attribution maps for $\Omega_m$, with Integrated Gradients values normalized between 0 and 1 (the black-and-white scale is chosen for optimal visualization of patterns).

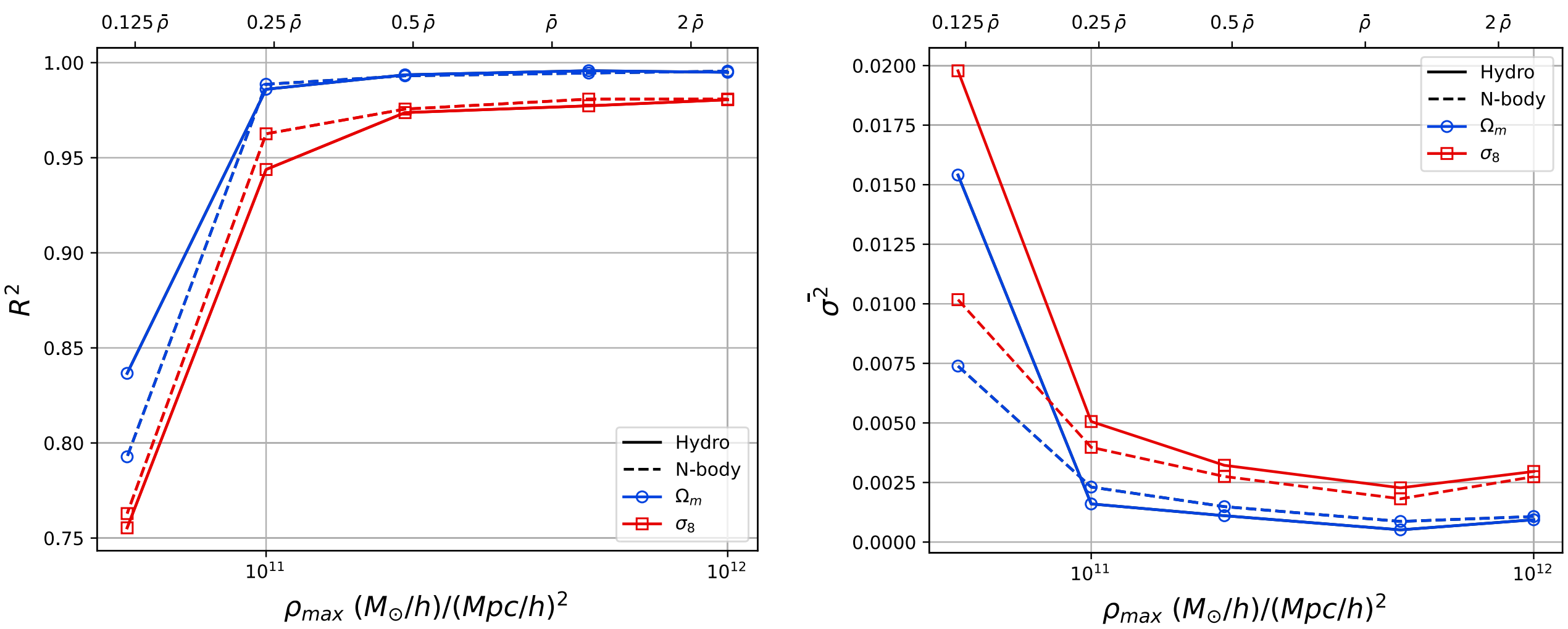

**Figure 10.** Overdensity cut inference. The $R^2$ scores (left) and average variance $\bar{\sigma}^2$ (right) of the predicted cosmological parameters $\Omega_m$ (circles in blue) and $\sigma_8$ (squares in red) as a function of $\rho_{\rm max}$. The solid lines represent the maps from the hydrodynamic simulations, and the dashed lines represent the maps from the $N$-body simulations. At low $\rho_{\rm max}$ there is a sharp increase in the $R^2$ and a decrease in variance, but already with $\rho_{\rm max} \simeq 0.25\bar{\rho}$ the inference precision and accuracy are saturated, implying that much of the information is in underdense regions.

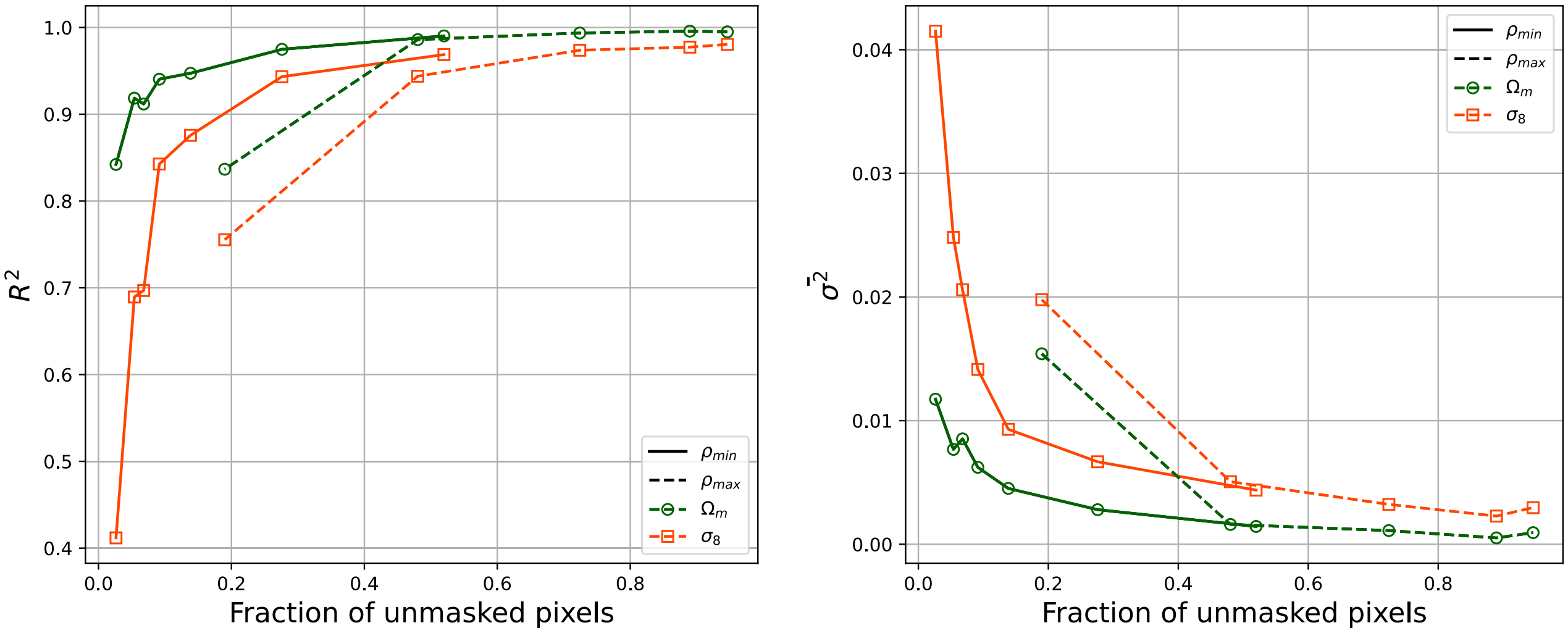

**Figure 11.** Random pixel augmentation inference. The $R^2$ scores (left) and average variance $\bar{\sigma}^2$ (right) of $\Omega_m$ (green circles) and $\sigma_8$ (orange squares) as a function of the fraction of unrandomized pixels for each of the $\rho_{\rm min}$ (solid lines) and $\rho_{\rm max}$ (dashed lines) cases for hydrodynamic maps. It can be seen for $R^2$ that the $\rho_{\rm min}$ lines are either above or equal to the corresponding $\rho_{\rm max}$ lines, implying that there is more information per pixel in overdense regions, particularly for $\sigma_8$.

### 3.3.3. Discussion of Density Cut Results

The minimum density cut analysis showed that removing the low-density regions of the maps gradually reduces the cosmological information content, with $R^2$ dropping by order 10% when $\rho_{\rm min} \sim 2\bar{\rho}$. Similarly, the maximum density cut analysis showed that the $R^2$ drops by order 10% when $\rho_{\rm max} \sim \bar{\rho}/4$. The fact that there is a sharp transition in $R^2$ at $\rho_{\rm max} \sim \bar{\rho}/4$ suggests that there is much information in voids, and the degradation in $R^2$ already at $\rho_{\rm min} \sim 2\bar{\rho}$ implies that there is much information at densities less than what corresponds to halos ($\rho \gtrsim 10^2\bar{\rho}$). This seems to suggest that halos provide comparatively less information than voids. However, it is also interesting to consider the information content per pixel to investigate whether the better performance of voids is simply a result of the fact that they have a larger volume and thus take up more pixels in the map compared to halos. Figure 11 shows the results of the minimum and maximum density cut analyses for the hydrodynamical simulations (from Figures 7 and 10), but instead of plotting the density cut value on the horizontal axis, we plot the fraction of pixels that remain after the density cut. Interestingly, we find that the $\rho_{\rm min}$ $R^2$ line is either above or equal to the corresponding $\rho_{\rm max}$ line for both $\Omega_m$ and $\sigma_8$, implying that there is more information per pixel in overdense regions compared to underdense regions, particularly for $\sigma_8$. This suggests that much of the information in underdense, void-like regions may be coming from their larger physical size and thus the large number of pixels that they span (their volume or area filling fraction) and/or from the shape of voids (A. E. Bayer et al. 2024).

## 4. Conclusions

Previous work by F. Villaescusa-Navarro et al. (2021b) has shown that CNNs can be used to extract cosmological information from 2D maps of simulated universes with astrophysical processes from the CAMELS project. The authors demonstrated that 2D maps of the total matter density field are particularly sensitive to the cosmological parameters $\Omega_m$ and $\sigma_8$, even when astrophysical processes are marginalized. In this work, we have investigated and interpreted the origins of the cosmological information extracted by the CNN from these maps.

We first considered Saliency Maps, Integrated Gradients, and GradentSHAP to understand which parts of the cosmic web CNNs are most sensitive to. We then investigated the information content as a function of Fourier scale $k$ cuts and density $\rho$ cuts—by randomly assigning values to pixels outside the cut thresholds—finding the following key results:

1. The network extracts cosmological information from deep voids in the cosmic web, as well as the topological structures of the filaments and halos. This suggests that the network is extracting information from the most extreme overdense and underdense regions.
2. For $k_{\rm max}$ cuts we find that the network is able to extract information about cosmological parameters, while marginalizing over baryonic effects, with only a 10% degradation between $k_{\rm max} = 50\,h/{\rm Mpc}$ and $k_{\rm max} = 10\,h/{\rm Mpc}$. This implies that CNNs obtain precise and accurate predictions of cosmological parameters with relatively aggressive scale cuts.
3. For $\rho_{\rm min}$ cuts we find that the network learns much information from underdense regions, with the inference quality dropping significantly for $\rho_{\rm min} \gtrsim \bar{\rho}$.
4. For $\rho_{\rm max}$ cuts we find that the network retains accuracy for $\rho_{\rm max} \gtrsim \bar{\rho}$, implying that the network is able to learn cosmological information using information from underdense regions alone.
5. Taking density cuts provides more information than randomly cutting regions of the map, implying that there is information in the spatially coherent morphological structure of the cosmic web, such as voids and filaments.
6. By comparing the results for hydrodynamic simulations to gravity-only $N$-body simulations, we find the quality of inference of $\Omega_m$ is the same with and without the modeling of astrophysical processes, while there is a slight degradation from astrophysical effects on $\sigma_8$. This implies that the network is actually learning from the morphological features of the cold dark matter field, even in the presence of baryonic effects, as if there were no baryons present.

7. While we find CNNs to be particularly sensitive to underdense regions—voids—we find that the information per pixel is actually larger in overdense regions. This implies that information from voids is partly due to their large size and/or their overall shape and structure (A. E. Bayer et al. 2024).

Our work confirms the complementarity of overdense and underdense regions in constraining cosmology (A. E. Bayer et al. 2021; C. D. Kreisch et al. 2021; M. Golshan & A. E. Bayer 2024). We note that while our results consider the fundamental information content in noise-free simulations, real observations will contain noise, which has been shown to more strongly degrade information in underdense regions compared to overdense regions (J. M. Zorrilla Matilla et al. 2020; M. Golshan & A. E. Bayer 2024).

Profoundly, our findings imply that there is little degradation in cosmological constraining power when performing inference in the presence of complex baryonic processes and when taking rather aggressive $k_{\rm max}$ or $\rho_{\rm max}$ cuts. Performing such cuts—essentially removing regions of the map that are most strongly affected by baryons (A. Amon & G. Efstathiou 2022)—would thus be a powerful way to obtain constraints that are robust to astrophysical effects that are difficult to model, such as the details of halo finders and different subgrid models. This is of great relevance for upcoming astrophysical surveys that measure observables such as galaxy clustering, weak lensing, 21 cm, and the Sunyaev–Zel'dovich effect, all of which are sensitive to the interplay of baryons and cold dark matter and often have to cut much data to ensure accurate cosmological inference.

Our results motivate the investigation of optimal data filtering methods. Typical cosmological analyses simply cut out data as a function of $k$, while here we have shown the benefits of $\rho$ cuts. Previous work has investigated the effect of $\rho$ cuts in the context of the power spectrum and bispectrum (M. C. Neyrinck et al. 2009; F. Simpson et al. 2011; A. Repp & I. Szapudi 2021). It would be worthwhile future work to look for a more optimal choice of filtering based on some combination of $k$, $\rho$, and other features of the data to enable robust inference at the field level. Moreover, while we have shown robustness to baryonic physics for a given subgrid model, it would be fruitful further work to further investigate different hydrodynamical simulation codes from CAMELS and to design filtering strategies to ensure robustness across different subgrid models.

## Acknowledgments

The convolutional neural networks have been trained using GPUs from the Rusty cluster at the Flatiron Institute. The authors thank Chaitanya Chawak and Natali de Santi for helpful conversations. The work of A.L. is supported by the TITAN ERA Chair project (contract No. 101086741) within the Horizon Europe Framework Program of the European Commission. The work of F.V.-N. is supported by the Simons Foundation. The CAMELS project is supported by the Simons Foundation. Details about the CAMELS simulations can be found at https://www.camel-simulations.org.

## Appendix
## Additional Tests and Clarifications

### A.1. Cross-testing between N-body and Hydrodynamic Simulations

We performed cross-application experiments to test whether networks trained on hydrodynamical IllustrisTNG maps generalize consistently to gravity-only ($N$-body) maps and vice versa, as a function of the overdensity ($\rho_{\rm max}$) and underdensity ($\rho_{\rm min}$) cuts.

For the $\rho_{\rm max}$ case, when the neural network is trained on the hydrodynamic maps and evaluated on $N$-body, both $\Omega_m$ and $\sigma_8$ are predicted with high accuracy ($R^2 \gtrsim 0.9$). This indicates that the network learns morphological features of the cold dark matter field while marginalizing over baryonic effects present in TNG. Since baryons modulate but do not fundamentally alter the large-scale morphology of the density field, the mapping learned in the presence of baryons remains valid when applied to $N$-body maps. The uncertainties are, however, somewhat larger in this case, consistent with the additional variance introduced by baryonic processes that are effectively marginalized over.

In contrast, when the network is trained on $N$-body and tested on IllustrisTNG, we observe a marked asymmetry. Predictions of $\Omega_m$ remain robust, as they are primarily constrained by large-scale structure that is less sensitive to baryonic physics. However, predictions of $\sigma_8$ degrade significantly, with $R^2$ values dropping close to zero. This arises because $\sigma_8$ is sensitive to the amplitude of small-scale fluctuations, which are suppressed in hydrodynamical runs owing to feedback processes such as supernovae and AGNs. A network trained exclusively on $N$-body data implicitly assumes that no such suppression exists and therefore misinterprets baryon-induced suppression as a shift in $\sigma_8$. This bias demonstrates that baryonic physics plays a central role in shaping $\sigma_8$ inference.

For the $\rho_{\rm min}$ case, the behavior is markedly different. The trends between hydro and $N$-body simulations are neither as obvious nor as symmetric. Within underdense regions (cosmic voids), baryonic physics leaves a much stronger imprint, and the discrepancies between hydrodynamical and $N$-body runs become significant. This leads to substantial information loss when cuts are applied in the $\rho_{\rm min}$ regime. In other words, void environments exhibit a breakdown of cross-simulation generalization, highlighting the sensitivity of low-density regions to baryonic processes.

This is in line with results from F. Villaescusa-Navarro et al. (2021b) that the difference between the astrophysical models of the training data set and the test data set leads to non-robust constraints on the cosmology. However, the results from the cross-testing case for overdensity cuts validate the significance of our results, as in the situation where the lack of baryonic effects does not significantly affect the cold dark matter distribution in features of the cosmic web the cross-tested model is indeed learning to infer the cosmology from the morphological features of the cosmic web.

### A.2. Impact of Simulation Volume on $\sigma_8$ Inference

The $\sigma_8$ parameter characterizes fluctuations on $8\,h^{-1}$ Mpc scales, raising the question of whether a $25\,h^{-1}$ Mpc simulation box contains sufficient information for robust inference. Although the number of independent long-wavelength modes is small in such a box, our inference does not rely exclusively on large-scale power but rather on the full morphology of the cosmic web. As shown in Figure 12, most of the constraining power saturates by $k \approx 20\,h\,\mathrm{Mpc}^{-1}$, corresponding to structures of size $\sim 0.31\,h^{-1}$ Mpc. The network gains substantial information from scales between $k \approx 5\,h\,\mathrm{Mpc}^{-1}$ (size $\sim 1.25\,h^{-1}$ Mpc) and $k \approx 20\,h\,\mathrm{Mpc}^{-1}$. A $25\,h^{-1}$ Mpc box contains $\mathcal{O}(20)$ independent features of size $1.25\,h^{-1}$ Mpc and $\mathcal{O}(80)$ of size $0.31\,h^{-1}$ Mpc, providing the CNN with ample independent structures to learn from. This enables the network to achieve high predictive accuracy ($R^2 \gtrsim 0.95$ for $\sigma_8$) even in a small-volume setting.

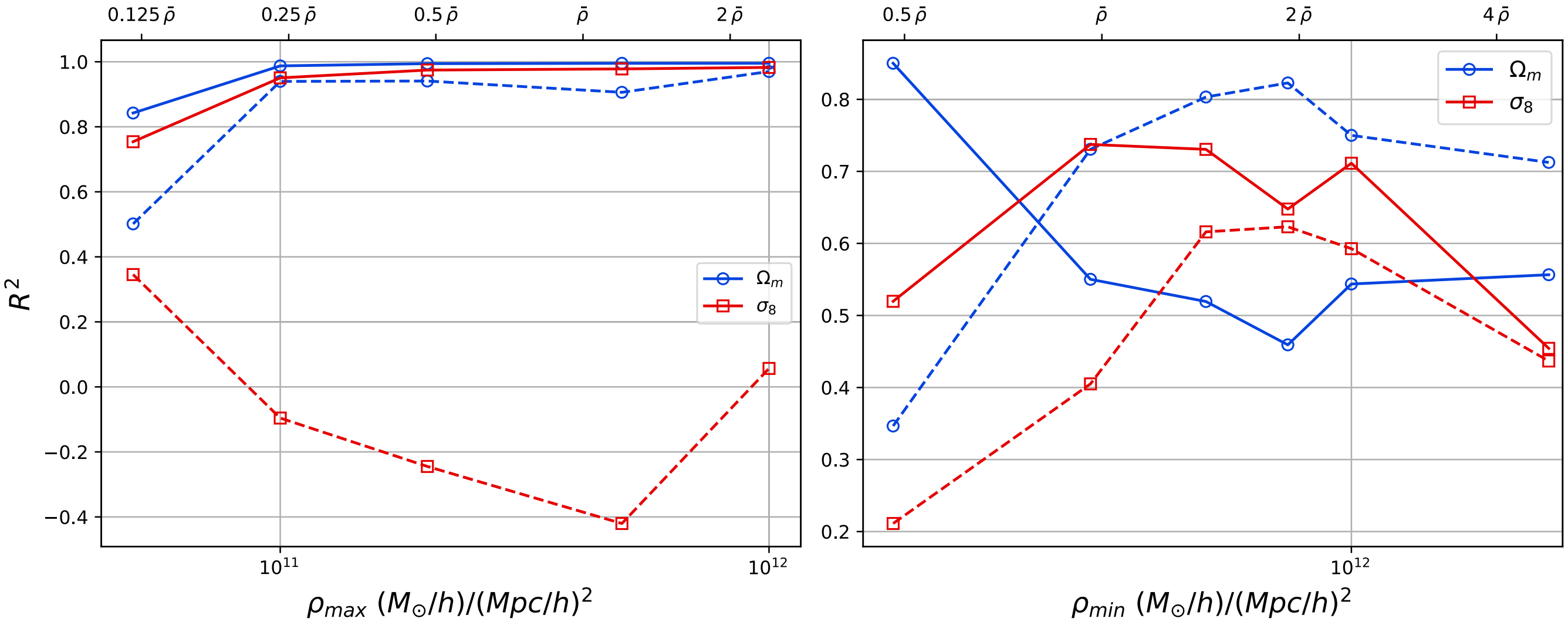

**Figure 12.** The variation in $R^2$ values for cross-testing case studies conducted for overdensity and underdensity cuts.

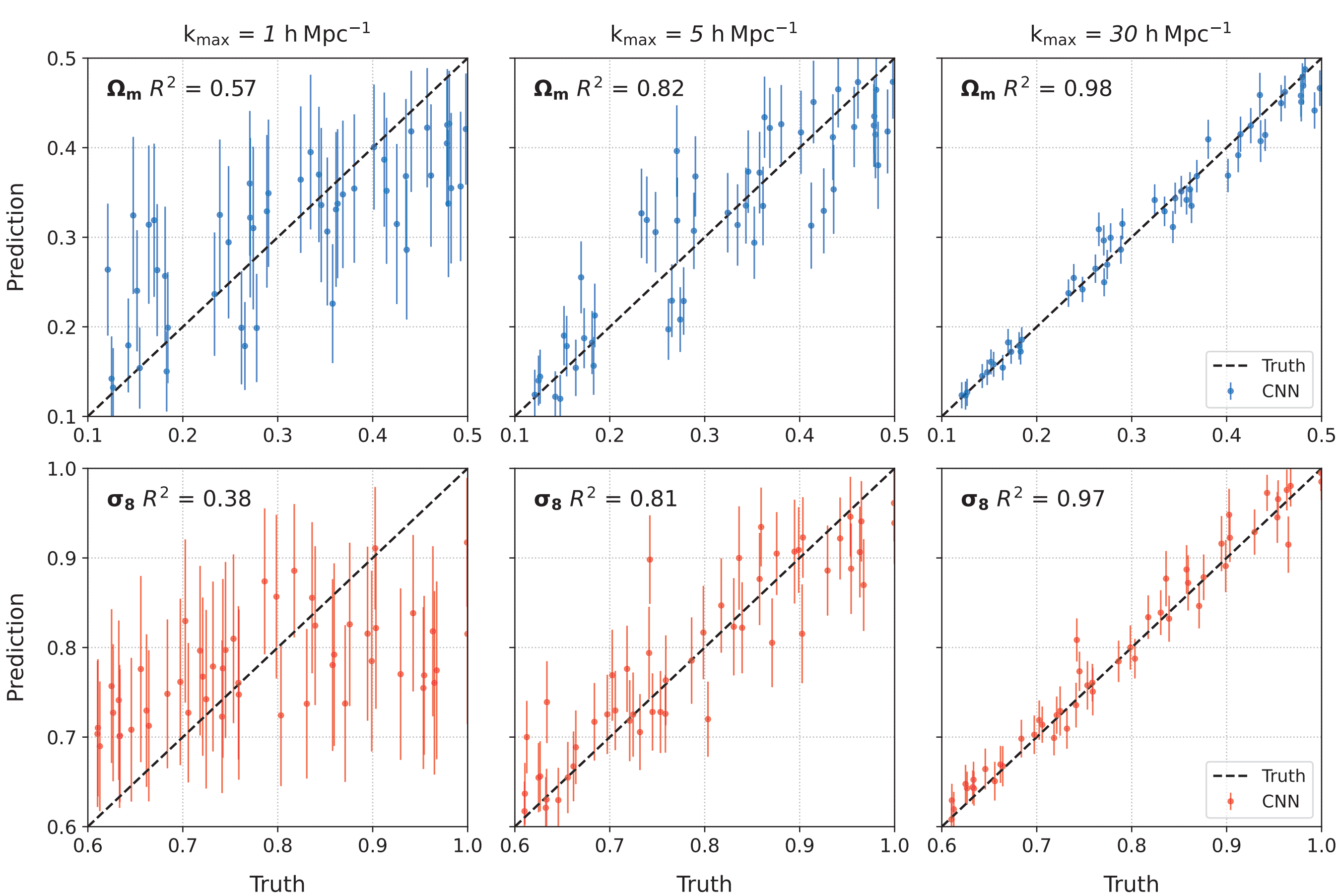

**Figure 13.** Scatter plots of predicted vs. true $\Omega_m$ and $\sigma_8$ for representative $R^2$ values corresponding to selected density cuts. High $R^2$ values correspond to points closely aligned along the diagonal, while lower $R^2$ values show increased scatter.

While very large scale modes ($\lambda \gtrsim 50\,h^{-1}$ Mpc) are absent from the CAMELS boxes, our results demonstrate that local morphological features such as the shapes of voids, halos, and filaments contain sufficient cosmological information for precise inference. This is consistent with the view that field-level inference derives much of its constraining power from non-Gaussian statistics on small to intermediate scales, rather than solely from the linear power spectrum on large scales. Extending this approach to larger-volume simulations (e.g., Millennium or Quijote) by subdividing them into 25 $h^{-1}$ Mpc patches would provide a direct test of generalization across volumes and constitutes an interesting avenue for future work.

### A.3. Interpretation of $R^2$ Values

To provide a more intuitive understanding of the $R^2$ values shown in Figure 12, we present scatter plots of predicted versus true cosmological parameters for selected representative $R^2$ thresholds (Figure 13). Each panel corresponds to a specific top-hat-$k$ cut at a particular $k_{\max}$ (as an example) with a characteristic $R^2$ score, illustrating the visual variability of the predictive performance with respect to $R^2$.

The scatter plots demonstrate that higher $R^2$ values correspond to tighter clustering of points along the best-fit diagonal, indicating more accurate predictions. Conversely, lower $R^2$ values exhibit increased scatter and systematic deviations, reflecting reduced information content in the particular wavenumber regime.

These plots complement the quantitative $R^2$ curves in Figures 7, 10, and 12, by providing a direct visualization of the predictive performance, helping to interpret the practical significance of the variation in $R^2$ across different analysis cases.

## ORCID iDs

Arnab Lahiry https://orcid.org/0009-0000-5950-1473
Adrian E. Bayer https://orcid.org/0000-0002-3568-3900
Francisco Villaescusa-Navarro https://orcid.org/0000-0002-4816-0455